\newread\testifexists
\def\GetIfExists #1 {\immediate\openin\testifexists=#1
    \ifeof\testifexists\immediate\closein\testifexists\else
    \immediate\closein\testifexists\input #1\fi}
\def\epsffile#1{Figure: #1}     

\immediate\openin\testifexists=e
    \ifeof\testifexists\immediate\closein\testifexists\else
    \immediate\closein\testifexists\input e\fipsf

\magnification= \magstephalf \tolerance=1600
\parskip=5pt
\baselineskip= 4.5 true mm \mathsurround=1pt
 \hsize=5.3 in
 \vsize=7.6 in
\font\smallrm=cmr8  
\font\medrm=cmr9  
\font\bigbf=cmbx12 \font\bigit=cmti12
    \def\Bbb#1{\setbox0=\hbox{$\tt #1$}  \copy0\kern-\wd0\kern .1em\copy0}
    \immediate\openin\testifexists=a
    \ifeof\testifexists\immediate\closein\testifexists\else
    \immediate\closein\testifexists\input a\fimssym.def 
\def\secbreak{\vskip 10pt plus .5in \penalty-200\vskip 0pt plus -.3in}
   \def\newsect#1{\secbreak\noindent{\bf #1}\medskip}
\def\hugeskip{\vskip12mm plus 3mm}
\def\Narrower{\par\narrower\noindent}   
\def\Endnarrower{\par\leftskip=0pt \rightskip=0pt}
\def\br{\hfil\break}    \def\ra{\rightarrow}
\def\a{\alpha}         \def\g{\gamma}  
\def\d{\delta}        \def\e{\varepsilon}
               
\def\m{\mu}         \def\f{\phi}            
\def\n{\nu}         \def\j{\psi}    
\def\r{\varrho}     \def\s{\sigma}  
        \def\th{\theta}  
              
  \def\W{\Omega}

\def\DD{{\cal D}}
 \def\TT{{\cal T}}
\def\cl{\centerline}    
\def\ni{\noindent}      \def\pa{\partial}   \def\dd{{\rm d}}
\def\tl{\tilde}         \def\bra{\langle}   \def\ket{\rangle}

\def\fn#1{\ifcase\noteno\def\fnchr{*}\or\def\fnchr{\dagger}\or\def
    \fnchr{\ddagger}\or\def\fnchr{\medrm\S}\or\def\fnchr{\|}\or\def
    \fnchr{\medrm\P}\fi\footnote{$^{\fnchr}$}
    {\scrunch#1\toe}\ifnum\noteno>4\global\advance\noteno by-6\fi
    \global\advance\noteno by 1}
    \def\scrunch{\baselineskip=11 pt \medrm}
    \def\toe{\vphantom{$p_\big($}}
    \newcount\noteno

\def\ffract#1#2{{\textstyle{#1\over#2}}}
\def\fract#1#2{\raise .35 em\hbox{$\scriptstyle#1$}\kern-.25em/
    \kern-.2em\lower .22 em \hbox{$\scriptstyle#2$}}

\def\half{\ffract12} 

\def\part#1#2{{\partial#1\over\partial#2}}
 \def\ref#1{${\vphantom{)}}^#1$}
\def\ex#1{e^{\textstyle#1}}

\def\bbf#1{\setbox0=\hbox{$#1$} \kern-.025em\copy0\kern-\wd0
    \kern.05em\copy0\kern-\wd0 \kern-.025em\raise.0433em\box0}

\def\low#1{{\vphantom{]}}_{#1}}
  
\def\ref#1{${\,}^{\hbox{\smallrm #1}}$}

\def\Gbar{\raise.13em\hbox{--}\kern-.35em G}
\def\lap{\setbox0=\hbox{$<$}\,\raise .25em\copy0\kern-\wd0\lower.25em\hbox{$\sim$}\,}
\def\glt{\setbox0=\hbox{$>$}\,\raise .25em\copy0\kern-\wd0\lower.25em\hbox{$<$}\,}
\def\gap{\setbox0=\hbox{$>$}\,\raise .25em\copy0\kern-\wd0\lower.25em\hbox{$\sim$}\,}

 \def\g44{\Big(1-{2M\over r}\Big)}

\def\in{\low{\rm in}}    \def\out{\low{\rm out}} \def\BH{\low{\rm BH}}

 {\ }\vglue 1truecm
 \rightline{SPIN-2000/06}\rightline{hep-th/0003004} \hugeskip
\cl{\bigbf The Holographic Principle
}\smallskip\cl{\bigit Opening Lecture} \hugeskip \cl{Gerard 't
Hooft }
\bigskip
\cl{Institute for Theoretical Physics}
\cl{University of Utrecht, Princetonplein 5}
\cl{3584 CC Utrecht, the Netherlands}
\smallskip
\cl{and}
\smallskip
\cl{Spinoza Institute}
\cl{Postbox 80.195}
\cl{3508 TD Utrecht, the Netherlands}
\smallskip\cl{e-mail: \tt g.thooft@phys.uu.nl}
\cl{internet: \tt http://www.phys.uu.nl/\~{}thooft/ } \hugeskip
\ni{\bf Abstract}\Narrower After a pedagogical overview of the
present status of High-Energy Physics, some problems concerning
physics at the Planck scale are formulated, and an introduction is
given to a notion that became known as ``the holographic
principle" in Planck scale physics, which is arrived at by
studying quantum mechanical features of black holes. \Endnarrower
\bigskip
\newsect{1. Introduction.}
To open an International School at which many important issues of
modern elementary particle physics will be discussed, it seems
appropriate to start with a bird's eye view of the recent
developments in the field. Another motivation to do so is that
fundamental physics today appears to have reached a new stage, at
which some reflection is needed over the past, in order to explain
our present standpoints and views, and to justify the kinds of
questions that we think we have to ask today, in order to enable
us to proceed in our field.

Before the '70s, we had the following picture of the fundamental forces. First, there was
{\it Quantum Electrodynamics} (QED), a very successful scheme to describe (nearly) all electric
and magnetic features of our beloved particles. It was understood how to perform impressively
accurate calculations by using perturbation expansions with respect to $\a=e^2/4\pi\hbar c\approx1/137$,
a small parameter\ref1. Although it was understood how to {\it renormalise\/} the apparent divergences
of the theory, it was still something of a mystery why this procedure worked at all, and indeed,
sometimes (for instance when electromagnetic mass differences were to be calculated) it did not
appear to work.

As for the {\it weak forces}, we only had an `effective' expression for the interaction\ref2, that however
was not renormalisable, which made it impossible to calculate any of the higher order radiative corrections.

The {\it strong force\/} was in an even worse state. It was
usually treated as a `black box', for which only the symmetry
pattern was well established. Several simplistic but quite
instructive models could be written down (the Gell-Mann-L\'evy
model\ref3, the dual resonance model\ref4), but they were mutually
incompatible, and since the coupling strength was large,
perturbation expansions, even if you could renormalise, appeared
to be meaningless.

Then, the revolution of the '70s came. The discovery that non-Abelian gauge theories are renormalisable\ref5
enabled us to make a very important step. We could now ask the question: ``What is the {\it most general\/}
perturbatively renormalisable quantum field theory?"\ref6

The answer turned out to be that our theory must consist of three kinds of basic particles, to be
represented by fundamental fields. They are distinguished by the value $S$ of the intrinsic spin:
\itemitem{$S=1$:} These particles cannot be described unless you have a (Abelian or non-Abelian) gauge theory.
The gauge group may be any local, compact Lie group $G$, for instance $G=SU(3)\otimes SU(2)\otimes U(1)
\otimes\dots$.
\itemitem{$S=\half$:} These particles must be described by Dirac fields, $\j^L$ and $\j^R$, of which $\j^L$
transforms as a $\bf 2\times1$ representation of the algebra
$SU(2)^{\rm Left}\otimes SU(2)^{\rm Right}$ of the Lorentz group
(in Euclidean notation) and $\j^R$ transforms as a $\bf 1\times2$.
Each of these fields may be in any kind of finite-dimensional
representation of the gauge group $G$, but there is a very
important restriction: an {\it anomaly\/} may arise in the
contributions of triangle diagrams to the matrix elements of axial
currents. It is not allowed to have currents with anomalies in
them, coupled to gauge fields. If $\j^L$ and $\j^R$ are in
different representations of the gauge group $G$, one must require
that their contributions to the axial anomalies nevertheless
cancel out\ref7. In the current version of the Standard Model,
this is indeed the case.
\itemitem{$S=0$:} Any set of scalar fields $\f$ may be present, in any finite representation of $G$.
Its self-interactions must be polynomial of degree four, and its
interaction with the fermions must be through gauge-invariant
Yukawa terms of the form ${\overline\j}\vphantom\j^L\f\j^R+$h.c.
Via the Higgs mechanism\ref8, these fields may produce masses for
vector particles and Dirac particles.\ref9\smallskip The present
Standard Model obeys all of these requirements --- and more: the
renormalization group equations tell us that the coupling
strengths vary as the energies increase, in such a way that
perturbation expansions can be applied up to extremely high
energies\fn{An uncertain factor here is the Higgs self-coupling,
since the Higgs mass is still unknown.\ref{10}}. Yet many
questions are still not answered:
\item{---} What determined Nature's choice for the gauge group $G$, the fermionic and scalar representations
of $G$, the number of leptonic and quark generations, and the values of the coupling strengths,
in particular the details of the Kobayashi-Maskawa matrix?\ref{11}
\item{---} How do we explain the large hierarchy of scales in Nature? The disparity between the
Planck scale, the electro-weak scale, and the scale(s) of the neutrino masses is just one of several
questions of this sort.
\item{---} What is the role of supersymmetry, and how is supersymmetry broken?\ref{12}
\item{---} How do we couple the gravitational force, and why is the cosmological constant
as small as it is, or if it vanishes altogether, again, why? There is no known symmetry that
`protects' the cosmological constant against renormalisation effects.\smallskip
\ni Different answers to some or all of these questions are presently being investigated. Judging
from past experiences, it must be of extreme importance to ask the right questions.

Are there any further useful results to be expected from
experiments? Three classes of experimental avenues have not yet
been completed, and may give us great improvements in our
understanding, although all of these are becoming more and more
difficult, demanding increasing skills of the experimenters:

\ni $a$) At increasing, higher energies, the following is to be expected, and I think will be done:
\item{---} The Higgs is there to be discovered.
\item{---} Supersymmetry partners of all presently known particles may be detected, hopefully
some time soon. At first sight, the fact that supersymmetric patterns were discovered in nuclear
physics\ref{13} has little to do with the question of supersymmetry among elementary particles, but it
may indicate that, as the spectrum of particles is getting more and more complex, some
supersymmetric patterns might easily arise, even if there is no `fundamental' reason for their
existence.
\item{---} Other new structures may also be found at higher energies. The most pleasant surprises will be
the unexpected ones, which may open up new fields. It is generally
believed that the present model will break down beyond a TeV or
so, and this energy level will be within reach in a decade or
so\ref{14}.
\smallskip\ni  These high energy experiments address the unknown physics in a direct manner, and
they are therefore most important.

$b$) On rare occasions, new results may also be expected from
precision experiments at lower energies. There have been a number
of interesting examples in the recent past:
\item{---} Atomic parity violations could be measured with better than one percent precision, in spite of the
fact that these are minute effects\ref{15}, yielding independent
confirmation of the effects due to $W$ and $Z$ exchanges within
atomic nuclei.
\item{---} The $K_L/K_S$ system is a beautiful laboratory. Precision measurements can be made of
the parameter $\e'/\e$, which may reveal features from deeply
inside, or possibly beyond, the Standard Model, as we will learn
at this School\ref{16}.
\item{---} Other known fundamental principles of Theoretical Physics can be put to a test,
such as $CPT$ invariance, relativity tests, the ratios $Q/M$ can
be compared between particles and antiparticles\ref{17}, the
Quantum Mechanics of gravitating systems can be investigated, etc.
\item{---} New ideas were launched suggesting that Newton's law of the gravitational force
might change at scales below a mm. This can be experimentally
tested.\ref{18}
\item{---} Tiny mass terms that produce mixing between various neutrino species can be
detected in dedicated experiments.\ref{19}
\item{---} And there are doubtlessly many more subtle effects that may be discovered and that
will alter our views concerning the fundamental
interactions.\smallskip \ni $c$) A third source of information is
cosmology. It used to be well within the domain of
Science-Fiction, but nowadays cosmological models are becoming
more mature. They yield precise predictions that can be verified
by astrophysical observations. Models of the inflationary universe
probe deeply into regions at extremely high energy, and so the
information they deliver is unique:
\item{---} Structures in the spectrum of the cosmic background radiation are predicted and
more detailed observations are to be expected.
\item{---} The distribution of galaxies is speculated to be due to quantum fluctuations in
a very early universe. They will be calculated and compared to what is observed.
\item{---} The search for dark matter continues. The outcome will deeply affect our thinking
about the fundamental interactions.
\item{---} Statistical analysis of distant galaxies may finally also reveal the presence of
a cosmological constant term in the Einstein-Hilbert action.
\item{---} Other tests of the models, for instance the baryon-antibaryon asymmetry and
$CP$ violation.

\ni In spite of this long list, there are reasons to worry about the increasingly difficult barriers
from behind which we are trying to understand the small-distance structure of our world.
Which purely theoretical approaches will help us find the answers? We have to concentrate on {\it fundamental
inconsistencies\/} in our present picture. There are many of these:
\item{---} As it was already mentioned, the hierarchies seen in the distance scales are not properly
explained by what is presently known.
\item{---} The apparent absence of a cosmological constant is at odds with what we understand
about Quantum Gravity.
\item{---} Indeed, quantizing gravity is still a deep problem. Superstring theory is vigorously
trying to bring the gravitational force under control, but it
surely is a wild animal. From superstrings came $D$-branes, from
$D$-branes came ``$M$-theory", but it has as yet not been possible
to even come close to an accurate formulation of the laws. These
ideas are of extreme importance, but new avenues must still be
found. \smallskip \ni What is known for sure is that Quantum
Mechanics works, that the gravitational force exists, and that
General Relativity works. The approach advocated by me during the
last decades is to consider in a direct way the problems that
arise when one tries to combine these theories, in particular the
problem of gravitational instability. These considerations have
now led to what is called ``the Holographic Principle", and it in
turn led to the more speculative idea of deterministic quantum
gravity. This theory, and the effects of {\it dissipation of
information}, will be discussed in a separate lecture. Our central
issue is: {\it What is Nature's bookkeeping system at the Planck
scale?}

\newsect{2. The inevitable existence of black holes}
In sufficiently large amounts of matter, gravitational collapse is
inevitable.\ref{20} There are various ways to derive this fact.
First, one may consider a stationary, spherically symmetric
configuration of matter, held together by gravity. Near the
surface, we assume that there is a region $r=r_1$ where the
temperature is low enough so that the density $\r_1$ there is
sufficiently high, say that of water. Assume that inside the
sphere $r=r_1$, there is a certain amount of mass $M_1$. The
pressure $p$ at this surface is still negligible. It is now easy
to argue that $M_1$ must be subject to an upper limit. In any
case, the pressure $p$ rises if we look at smaller distances $r$
from the centre. If the density were constant, and the general
relativistic effects negligible, then one could readily compute
the pressure at the centre. However, the density is likely to
increase if we go down. In fact, if we assume our material to be
{\it non-exotic}, then a finite compressibility follows. Matter is
defined to be non-exotic if the speed of sound $v_s$ is less than
the speed of light, $c$. This means that the gradient of the
gravitational field will rise, and hence the gradient in the
pressure will become steeper, and this will cause an instability.
If $M_1$ was chosen large enough, there will be a point
$r=r_2<r_1$ where the pressure becomes infinite. Even without any
other relativistic arguments, this gives us as a limit:
$\ffract23\pi G_N\r\low{1}r_1^2<1$. Adding general relativistic
effects correctly will give a more stringent limit, as I will
briefly explain later.

But first: does matter have to be non-exotic? Suppose the speed of
sound exceeds the speed of light. Would there be an immediate
contradiction? Special relativity would normally demand that no
signal can go faster than light. However, the reason why we demand
this is causality: no signal should be able to propagate {\it
backwards\/} in time. This is then combined with demanding
Lorentz-invariance. However, matter in equilibrium represents a
preferred Lorentz frame, so we could drop the latter demand.
Still, there must be restrictions. Consider two regions in which
matter has different local velocities. Imagine two adjacent pipes
in which matter streams in opposite directions, and in both, sound
goes faster than light, also in opposite directions. Due to the
time shifts when Lorentz transforming, an outside observer may see
both signals move backwards in time. This, in principle, could
then generate a closed loop of information transport in
space-time, which is an undesirable situation. This we must
forbid.

Even exotic matter, however, will not be able to stop black holes
from being formed. This is seen if we insert the complete general
relativistic equations instead of our above pseudo-relativistic
argument. These are the so-called Tolman-Oppenheimer-Volkoff
equations\ref{21} These equations tell us how density and pressure
increase when followed inwards, given some equation of state. It
is an elementary exercise to solve these equations for constant
density. Even then, one finds that, due to space-time curvature,
the pressure diverges to infinity at a finite radius, if we start
with too much mass and a too high density at a too small radius on
the surface.

We can however also produce a black hole from ordinary matter at
zero pressure. Consider a spherically symmetric arrangement of
matter in the form of a shell, with some finite thickness. We
allow the shell to contract due to its own gravitational field.
Inside the shell, there is no gravitational field at all,
something that one can understand using the same arguments that
tell us that inside a conducting metal sphere there is no electric
or magnetic field. If the original amount of material was big
enough, the contraction will proceed, and, in the limit of zero
pressure and purely radial, spherically symmetric motion, the
equations can easily be solved exactly. We obtain flat space-time
inside, and a pure Schwarzschild metric outside. As the ball
contracts, a moment will arrive when the Schwarzschild horizon
appears. From that moment on, an outside observer will no longer
detect any radiation from the shell, but a black hole instead.
\newsect{3. Hawking radiation and quantum states}
The standard generally relativistic black hole solution has as a
special feature that shortly after its formation, no signals will
be seen coming out. It should be truly black. As is well-known,
this picture changed when Hawking\ref{22} discovered an elementary
consequence of quantum field theory when applied to fields living
in the black hole metric. The rearrangement of creation and
annihilation operators is such that the states near the horizon
are not truly vacuum, but they contain a precisely computable
density of particles, which are emitted as black body radiation at
a temperature given by\ref{22} $$kT_H\ =\ {\hbar c^3\over 8\pi
G\,M_{\rm BH}}\ .\eqno(3.1)$$ This result allows us to compute the
{\it density of quantum states\/} of a black hole. The easiest way
to do this is by using thermodynamics. However, one could object
that a black hole is not truly in thermodynamic equilibrium; if
energy is added to a black hole, its mass and its size will
increase, and consequently its temperature will drop.

We can avoid thermodynamics by deriving the spectral density of a
black hole directly from its Hawking temperature. All one needs is
some form of time reversal invariance\ref{23}. We have at our
disposal both the {\it emission rate\/} (the Hawking radiation
intensity), and the {\it capture probability\/}, or the effective
cross section of the black hole for infalling matter.

In units at which $G=\hbar=c=1$, the cross section $\s$ is approximately:
$$\s\,\approx\,2\pi R^2\,=\,8\pi M^2\,,\eqno(3.2)$$
and slightly more for objects moving in slowly. The emission probability $W\dd
t$ for a given particle type, in a given quantum state, in a large volume
$V=L^3$ is:
$$W\dd t\,=\,{\s({\bf k}) v\over V}\,\ex{-E/kT}\dd t\,,\eqno(3.3)$$
where $\bf k$ is the wave number characterizing the quantum state, $v$ is the
particle velocity, and $E$ is its momentum.

Now we {\it assume\/} that the process is also governed by a Schr\"odinger
equation. This means that there exist quantum mechanical transition amplitudes,
$$\eqalign{\TT\in\,&=\,\BH\!\bra M+ E|\TT|M\ket\BH|E\ket\in \,,\cr{\rm and}\qquad
\TT\out\,&=\,\BH\!\bra M|\,\out\!\bra E|\TT|M+ E\ket\BH\,,}\eqno(3.4)$$
where the states $|M\ket\BH$ represent black hole states with mass $M $, and
the other states are energy eigenstates of particles in the volume $V$. In terms
of these amplitudes, using the so-called Fermi Golden Rule, the cross section
and the emission probabilities can be written as
$$\eqalignno{\s\,&=\,|\TT\in|^2\r(M+ E)/v\,,&(3.5)\cr
W\,&=\,|\TT\out|^2\r(M){1\over V}\,.&(3.6)\cr}$$
where $\r(M)$ stands for the level density of a black hole with mass $M$. The
factor $v^{-1}$ in Eq.~(3.5) is a kinematical factor, and the factor $V^{-1}$
in $W$ arises from the normalization of the wave function.

Now, time reversal invariance relates $\TT\in$ to $\TT\out$. To be precise, all
one needs is $PCT$ invariance, since the parity transformation $P$ and charge
conjugation $C$ have no effect on our calculation of $\s$.
Dividing the  expressions (3.5) and (3.6), and using (3.3), one finds:
$${\r(M+ E)\over\r(M)}\,=\,\ex{E/kT}\,=\,\ex{8\pi ME}\,.\eqno(3.7)$$
This is easy to integrate:
$$\r(M)\,=\,\ex{4\pi M^2 + C}\,=\,\ex S\,.\eqno(3.8)$$

We can rewrite this as $$\r(M)\,=\,2^{A/A_0}\,,\eqno(3.9)$$ where
$A$ is the horizon area and $A_0$ is a fundamental unit of area,
$$A_0\,=\,4 \ln 2\,L_{\rm Planck}^2.\eqno(3.10)$$ This suggests a
spin-like degree of freedom on all surface elements of size $A_0$,
see Fig.~1. \midinsert\cl{\epsffile{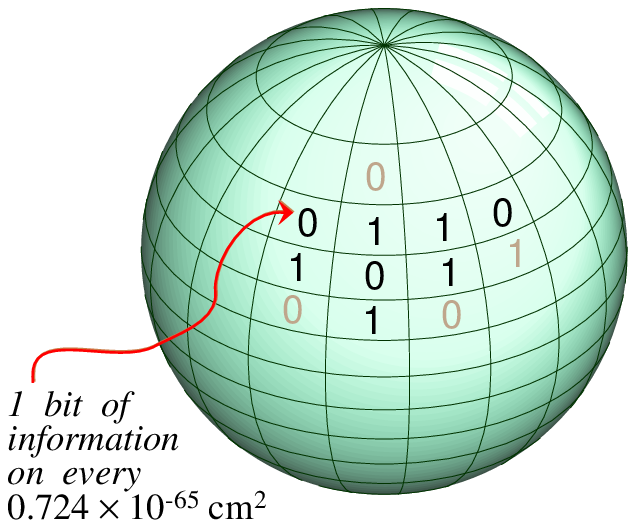}} \cl{Figure 1.
Information on a black hole horizon.}\bigskip
\endinsert
The importance of this derivation is the fact that the
expressions used as starting points are the {\it actual} Hawking emission rate
and the {\it actual} black hole absorption cross section. This implies that,
if in more detailed considerations divergences are found near the horizon,
these divergences should not be used as arguments to adjust the relation
between entropy and level density by large renormalization factors.

\newsect{4. The quantum information problem.}
It is tempting to conclude from the arguments presented above that
the `black hole states' form a natural extension of the spectrum
of elementary particles. The lightest particles are known and have
been identified as photons, neutrinos, electrons, muons, mesons,
baryons, and onwards to the heavy leptons, the Higgs and so forth.
The series could continue with as yet unknown particles in the
`desert' between 1~TeV and $10^{19}$~GeV, and beyond that region
the first superstring recurrences could exist. The `most pointlike
objects' beyond the Planck mass must undoubtedly be black holes,
simply because any sufficiently compact object with sufficiently
high mass must carry a gravitational field and a horizon
associated with that. Apparently, we now know the spectrum of the
objects in this range, apart from the unknown multiplicative
constant $e^C$ in Eq.~(3.8).

It should be possible to handle these objects just as all quantum objects
when we consider quantum mechanical amplitudes at high energies: they are
represented as propagators describing intermediate states. Theoretical Physics
should give us the computational rules, comparable to Feynman rules, for
computing these amplitudes. What have we got?

The behaviour of quantum fields near the horizon of a black hole follows
from the expression for $\dd s$, the infinitesimal invariant distance
element according to General Relativity:

$$\dd s^2=-\left(1-{2M\over r}\right)\dd t^2+{\dd r^2\over
1-2M/r}+r^2\dd\W^2\,, \eqno(4.1)$$ where $\dd\W^2$ stands for
$\dd\th^2+\sin^2\th\dd\f^2$. Writing $$r-2M\ =\ e^\s\ ,\qquad \dd
r= (r-2M)\dd\s\ ,\eqno(4.2)$$ we see that $$\dd
s^2=\left(1-{2M\over r}\right)\left(-\dd t^2+r^2\dd \s^2+{r^2\over
1-2M/r}\dd\W^2 \right)\,.\eqno(4.3)$$ At high energies, the
conformal factor has little effect on the wave equations for light
particle species. Consider a wave equation close to the horizon,
such as $$\pa_\m\sqrt{-g}\,g^{\m\n}\pa_\n\f\ -\ m^2\sqrt{-g}\,\f\
=\ 0\,.\eqno(4.4)$$ Since it contains the inverse of the metric
tensor $g_{\m\n}$, the contribution from the angular part in
Eq.~(4.3) becomes insignificant, and likewise the mass term. Thus,
the wave equations become 2-dimensional, and they generate plane
waves in the $\s$-$t$ direction as if $\s$-$t$ space were flat.
This means that one gets an unlimited number of oscillations near
the horizon, as $\s\ra-\infty$. Quite unlike the case for systems
such as the hydrogen atom, we see that the boundary condition at
the horizon is ill-defined. The quantum states can generate an
unlimited number of modes here. From this exercise, one would
conclude that the density of quantum states for a black hole is
not at all finite.

The physical origin of this divergence is not difficult to
identify. Particles may move into a black hole, but, as long as we
stick to linear field equations, the particles emerging from the
black hole cannot at all be related to the ingoing ones. There
cannot be any reflection against the horizon, since there should
be an infinite time delay. Here we see that the situation with
ingoing and outgoing spinors and vector particles will be equally
hopeless.

Two other conundrums are closely related to the problem just
signalled. First, we have the {\it quantum decoherence\/} problem.
This problem becomes apparent when the Hawking effect is
calculated explicitly. The initial state of elementary particles
before the formation of a black hole is described in terms of
various Fourier modes of their fields. All of these modes then are
associated to observable operators. In a Heisenberg picture of the
quantum states, these operators become time dependent. Part of the
Fourier modes of the initial fields now enter into the black hole,
and only the operators associated to the modes that emerge out of
the black hole correspond to observables at a later time. The
expectation values of these late-time observables turn out to be
described by a thermal density matrix. In terms of the basis of
states generated by the late-time observables alone, this density
matrix turns out to have eigenvalues less than one, which is
characteristic for a not fully coherent quantum state. This
situation is similar to what one gets in a condensed matter system
if one allows observable particles to escape, and subsequently
omits the quantum states that they represent.

In the case of a black hole, the missing particles are the absorbed ones. If we were
forced to keep these particles in our quantum description, an even worse infinity of
quantum states would result.

A description of the `information problem' that is easier to
understand is the following. Choose a coordinate frame in which
the formation of a black hole looks more or less regular. Ingoing
particles are then seen to enter at rather late times. If now the
returning particles were assumed to be not totally independent of
the ingoing ones, one would have to accept the observation that,
somehow, the information contained in the ingoing particles has
been transferred to the outgoing ones. The outgoing ones, however,
all belong to the Fourier modes that arose as quantum oscillations
at the point where the black hole was formed, way back in the
past. How could the required information have been imprinted on
these particles, if they have already been there for such a long
time?

A mathematically impeccable observation was made by
Hawking\ref{24}: the black hole space-time describes {\it two\/}
universes, not one, and these two are connected by a `wormhole'.
After a black hole is formed, the quantum wave functions of
elementary particles spread over these two universes, and they
become intertwined. Cutting off the information concerning the
contents of the `hidden' universe will leave the other universe in
a quantum mechanically decoherent state.

From a physical point of view, however, this argument is unsatisfactory. It implies that
black holes are fundamentally different from all other forms of matter in the sense that
they appear to produce decoherence. In all respects this result is equivalent to saying that
the scattering matrix elements involving black holes {\it are not fixed by our theory}, but
carry an uncertainty, distributed in some well-defined way. So, what we really have here
is an `uncertain theory'. Our theory is incomplete. We should not be satisfied with that.
Perhaps new physics can remove this uncertainty.

\newsect{5. The Scattering Matrix Ansatz.}
How do we `improve' our theory? Naturally, one may think of
including more interactions; obviously, the procedures applied
thus far assumed ingoing and outgoing fields not to interact ---
Eq.~(4.4) is after all linear in $\f$. At first sight it seems
that including interactions will resolve the paradox. As $\s$
approaches $-\infty$, the plane $\f$ waves enter conformally into
ever smaller regions of space and time. Effectively, the
gravitational couplings increase rapidly, and as $e^\s$ approaches
the Planck length, this effective coupling becomes super strong.
An alternative way to verify this is by switching towards a
coordinate frame that is locally regular near the horizon. Such a
frame is given, e.g., by the Kruskal coordinates $\{x,\,y\}$:
$$xy=\left({r\over2M}-1\right)e^{r/2M}\ ;\quad
x/y=e^{t/2M}\,.\eqno(5.1)$$ Writing $$x^0=x-y\ ;\quad
x^1=x+y\,,\eqno(5.2)$$ one finds the metric to be regular near
$x\approx y\approx 0$. Particles sent in in the far past will
align close to the axis $x=0$, and particles going out in the
distant future align close to $y=0$. A boost in the Schwarzschild
time parameter $t$ corresponds to a Lorentz transformation in
$(x,\,y)$ space, where the scale is set by the mass parameter $M$.
If we consider time lapses long compared to $M$, the Lorentz
boosts separating ingoing and outgoing particles become
horrendous. Thus we see that ingoing and outgoing particles meet
each other near $x=y=0$ at tremendously large c.m. energies. Even
if we could neglect Standard Model interactions at these energies,
the gravitational interactions, which grow with the energy
squared, can no longer be ignored from some point onwards.

This observation however does not resolve the decoherence problem. Even with the interactions
in place, one may still argue that information is drained by the black hole, and a theory
for pure states interacting with pure states without decoherence does not follow. A more
powerful approach is wanted.

It is strongly advocated now to start from the other end: we must {\it assume\/} that there exists
a quantum mechanically fully coherent scattering matrix $S$. The assumption is somewhat dogmatic;
we cannot prove it from first principles, other than demanding the {\it existence\/} of a theory.
Even if standard techniques at best only provide us with some `distribution' for the physical scattering matrix
elements, we assume that the `true' scattering matrix elements are exactly defined. Even if no
theory would exist to derive them, they could in principle be derived from experiment.

Demanding consistency with existing theory however gives us
important constraints. Indeed, the scattering matrix can now
almost be derived from the information we already have. The
calculations have been presented at length elsewhere, so here we
give a summary.

The dominant interaction is assumed to be the gravitational one,
simply because the c.m. energies tend to infinity. Other
interactions, such as in particular the electro-magnetic one, can
be corrected for later (the effects from electro-magnetism are
important, but they do not affect the main structure that will be
obtained). The procedure then is as follows.\ref{25}

First, assume a black hole with some well-specified initial
history of {\it ingoing} particles, for instance we specify the
way in which a star imploded to give this black hole, and
afterwards more objects may have fallen in at later times. We
assume that all this leads to a {\it pure\/} quantum state, to be
referred to as the state $|1\ket$. It evolves and decays in some
prescribed way. It leads to some superposition of many possible
states for the outgoing particles, including states describing the
final explosion.

Now, we consider the same state $|1\ket$, but we either add or remove one ingoing particle,
and we call this state $|1,\,\d p\ket$, where $\d p$ stands for the momentum (and possible other
details) of the extra ingoing object. What can be done now is a calculation, as detailed as possible,
of the effects this extra ingoing particle has on the outgoing objects. Surely there is interaction.
The gravitational one is most interesting. It leads to a {\it shift\/} of the outgoing wave functions.
This shift is simply the Shapiro delay due to the gravitational field of the ingoing object. The
calculation is in principle entirely straightforward, but has to be done with some care since the
ingoing object goes essentially with the speed of light. The shift of the Kruskal $y$ coordinate
is found to be
$$\d y=p\in\cdot G(|\tl x\in-\tl x\out|)\,,\eqno(5.3)$$
where $G$ is a simple calculable function of the coordinates $\tl x$ on the horizon. In the limit where the
black hole is large and the separations on the horizon small, we can approximately view $\tl x$ as
flat coordinates, and in that limit, the function $G$ is proportional to $\log|\tl x\in-\tl x\out|$.
We then find that this shift obeys a Laplace equation:
$$\tl\pa^2\d y=-C\cdot p\in\cdot \d^2(\tl x\in-\tl x\out)\,.\eqno(5.4)$$
$G$ is therefore a Green function.
Because of this shift, the outgoing state $|\j\out\ket$ turns into
$$\exp\left(i\int\dd^2\tl x\,P^+\out(\tl x)\d y(\tl x)\right)|\j\out\ket\,.\eqno(5.5)$$
Writing $\d y(\tl x)=\int\dd^2\tl x'\, G(\tl x-\tl x')\,\d p\in(\tl x')$, the new state is
$$|\j'\out\ket=\left(\exp\,i\int\int\dd^2\tl x\,\dd^2\tl x'\,P^+\out(\tl x)G(\tl x-\tl x')
\d p\in(\tl x')\right)|\j\out\ket\,.\eqno(5.6)$$
And now we can repeat this many times. Let $P^-(\tl x')$ stand for the total momentum
(in Kruskal coordinates) for all particles added, or subtracted using a minus sign, from the
state $|1\ket$. Then we have
$$|\j\out\ket=\left(\exp\,i\int\int\dd^2\tl x\,\dd^2\tl x'\,P^+\out(\tl x)G(\tl x-\tl x')
P^-\in(\tl x')\right)|1\ket\,.\eqno(5.6)$$
This way, the state $|1\ket$ can be used as a universal reference state. The true state is then
specified by giving $P^-\in(\tl x)$.

After some simple manipulations, we find that both the initial and the final state could be
described by giving the transverse coordinates $\tl x^{(i)}$ and the radial momenta $p^{(i)}$
for all particles. We get
$$\eqalign{{}\out\bra& q^1,\tl y^1,\,q^2,\tl y^2,\,\dots||\,p^1,\tl x^1,\,p^2,\tl x^2,\,\dots\ket\in=\cr
&=\int\DD X^+(\tl x)\,\DD X^-(\tl x)\exp\bigg(\int\dd^2\tl x\big[-i\tl\pa X^+\tl\pa X^-+\cr
&+\sum_ii\d^2(\tl x-\tl x^i)p^iX^+(\tl x)-\sum_ii\d^2(\tl x-\tl y^i)q^iX^-(\tl x)\big]\bigg)\,,}
\eqno(5.7)$$
and if the in- and outgoing particles are described by transverse wave functions $e^{i\tl p^i
\tl x^i}$ and $e^{i\tl q^i\tl y^i}$, then another set of integrations has to be performed, over
the transverse coordinates $\tl x^i$ and $\tl y^i$. All of this yields an amplitude that is very much
reminiscent of a string amplitude, with the exception of the $i$ in front of the `kinetic' term
$(\pa X)^2$ in Eq.~(5.7). In Eq.~(5.7), Newton's constant $G_N$ has been normalized according to
$$8\pi G_N=1\,.\eqno(5.8)$$

\newsect{6. Fock space.}
The result of the previous section appears to be beautiful. We managed to construct the $S$-matrix
using only known facts about the gravitational interaction between fast moving objects. In addition,
it appears not to be too difficult to impose unitarity for this scattering matrix. Unitarity just fixes
the {\it measure\/} of the functional integration in (5.7). Only the phase then remains undetermined,
but it was arbitrary anyway since the amplitudes in question violate many of the conventional conservation
laws such as all combinations of baryon and lepton number.

There are, however, two problems, both having to do with the Hilbert space in terms of which this
scattering matrix appears to be defined.

\ni Problem \# 1: the space of all momenta, $\{P^\pm(\tl x)\}$, is
infinite dimensional, even for small black holes, whereas we
expected a finite total number of states (the entropy was supposed
to be finite). So, this Hilbert space is far too large.

\ni Problem \# 2: the space of all momentum distributions,
$\{P^\pm(\tl x)\}$, is far too {\it small\/} to accommodate for
all possible particle configurations. If two or more particles
enter at the same transverse point $\tl x$, then, in our
expressions, only the total momentum counts. States for which the
total momentum distributions are identical will be
indistinguishable, and since we want our scattering matrix to be
unitary, these states must be {\it identical}. This is not Fock
space for elementary particles as we are used to.

It is important to note, on the other hand, that string amplitudes, which are like Eq.~(5.7) but without
the $i$ in the kinetic term, share the same feature: states with two or more particles entering the
string world sheet at one point, are indistinguishable from states with just a single particle entering
at that point. Distinctions only come after the $\tl x$ integrations, at which the particle number
becomes unambiguous.

\newsect{7. The holographic principle.\ref{26}}
We have reached a point where, for a proper description of the
particle states in the vicinity of a black hole, a {\it
two\/}-dimensional function is required: the momentum distribution
over a two-dimensional coordinate on the horizon. In addition,
this function must be further reduced, since it must effectively
contain not more than one $\Bbb Z(2)$ variable per surface element
$A_0$ (see Eq.~(3.10)). A comparison with a holographic photograph
is quickly made. In a holographic set-up, a laser beam shines onto
some three-dimensional object, and the reflected light interferes
with an unperturbed laser beam. The interference pattern is
registered on a photographic plate. In turn, after having
developed the plate, we can shine a laser beam on it. An image of
the three-dimensional object re-emerges. This appears to be a way
to register three-dimensional objects on a two-dimensional
photographic plate.

Now imagine that the photographic plate has a limited resolution, and that its colouring can
only be black-and-white, no gray tones. In that case, the image we see of the original object
will be blurred somewhat, since information went astray. This must actually be the situation in
our description of particles entering a black hole: the momentum distribution cannot
represent as many details as a fully three-dimensional description: our image of the universe is
blurred. Of course, since it is the Planck scale where this limit is attained, in practice
we perceive our universe very sharply.

Although this holographic nature of our description of the
particles appears to apply only for particles entering a black
hole, one may argue that it must have a much more universal
validity. According to general relativity, there should exist a
direct mapping that relates physical phenomena in one setting
(with a gravitational field present) to another one (freely
falling coordinates). Normally, the mapping goes both ways. It is
indeed unlikely that freely falling particles can be described in
more detail than the limits set by the holographic principle: one
bit of information per surface element of size $A_0$. It can be
computed that the energy needed to detect more details would be so
large that gravitational collapse would be inevitable; the entire
scene would be absorbed by a black hole -- and indeed be
impossible to observe at all!

This is what we found out about Nature's book keeping system: the data can be written onto
a surface, and the pen with which the data are written has a finite size.

\newsect{References}
 \item{1.} see for instance B.E. Lautrup et al, {\it Phys. Reports}
{\bf 3C} (1972) 193.
\item{2.} R.E.~Marshak {\it et al\/}, {\it
Theory of Weak Interactions in Particle Physics}, Wiley Intersc.,
1969, SBN 471 57290 X; E.C.G. Sudarshan and R.E. Marshak, Proc.
Padua-Venice Conf.  on  ``Mesons
     and Recently discovered Particles", 1957, p. V~-~14,  reprinted in:
     P.K.~Kabir, ``Development of Weak Interaction  Theory",  Gordon  and
     Breach, 1963, p. 118;
E.C.G. Sudarshan and R.E. Marshak, {\it Phys.~Rev.} {\bf  109} (1958) 1860; \br
R.P. Feynman and M. Gell-Mann, {\it Phys.~Rev.} {\bf  109} (1958) 193
  \item{3.} M. Gell-Mann and M. L\'evy, {\it Nuovo Cim.} {\bf  16} (1960) 705;  \br
 B.W. Lee, {\it Nucl.~Phys.} {\bf B9} (1969) 649; J.-L. Gervais and B.W.  Lee, {\it Nucl.
      Phys. \bf B12} (1969) 627; K. Symanzik, {\it Lett. Nuovo  Cim. \bf   2}  (1969)  10,
      {\it Commun.~Math.~Phys.} {\bf 16} (1970) 48.\br
 B.W. Lee, ``Chiral Dynamics", Gordon and Breach, New York, 1972, ISBN 0 677 01380 9 (cloth);
    0 677 01385 X (paper)
\item{4.}S. Mandelstam, {\it Phys. Reports \bf 13c} (1974) 259;\br
 G.~Veneziano, {\it Nuovo Cimento \bf 57A} (1968) 190;\br
 J.H.~Schwarz, {\it Phys.~Reports \bf 8C} (1973) 269;\br
 V.~Alessandrini {\it et al}, {\it Phys.~Reports \bf 1C} (1971) 269.
\item{5.} G.~'t Hooft, {\it Nucl.~Phys.} {\bf B33} (1971) 173; {\it id},
 {\it Nucl.~Phys.} {\bf B35} (1971) 167 .
\item{6.}J. Cornwall, D. Levin and G. Tiktopoulos, {\it Phys.~Rev.~Lett., \bf 30}  (1973)
     1268, {\bf 31} (1973) 572; C.H. Llewellyn Smith, {\it Phys.~Lett. \bf B46} (1973) 233.
\item{7.} S.L.~Adler and W.A.~Bardeen,
 {\it Phys.~Rev.}~{\bf  182} (1969) 1517;  W.A.~ Bardeen,
      {\it Phys.~Rev.}~{\bf  184} (1969) 1848.
\item{8.}P.W. Higgs, {\it Phys. Lett \bf 12} (1964) 132; {\it Phys.~Rev. Lett. \bf   13 } (1964)  508;
     {\it Phys.~Rev.} {\bf  145} (1966) 1156.
\item{9.}F. Englert and R. Brout, {\it Phys.~Rev.~Lett.} {\bf 13} (1964) 321.
\item{10.}  See for instance: E.~Accomando {\it et al}, {\it Phys. Reports \bf 299} (1998) 1, and
P.M.~Zerwas, {\it Physics with an $e^+e^-$ Linear Collider at High Luminosity},
Carg\`ese lectures 1999, preprint DESY 99-178.
\item{11.} M.~Kobayashi and T.~Maskawa, {\it Progr.~Theor.~Phys. \bf 49} (1973) 652.
\item{12.} P. Fayet and S. Ferrara, {\it Phys. Reports} {\bf 32C} (1977)
249; \br S.~Ferrara, {\it Supersymmetry}, North Holland, World
Scientific, 1987, Vols 1 and 2.\br Supersymmetry has a vast
literature. See for instance the collection of papers in:
S.~Ferrara, {\it Supersymmetry}, Vol.~1, North Holland, Amsterdam,
etc., 1987.
\item{13.} F.~Iachello and A.~Arima, {\it The Interacting Boson Model}, Cambridge Univ. Press,
1987.
\item{14.} A.~Zichichi, {\it Subnuclear Physics, The First Fifty
Years. Highlights from Erice to ELN.} Eds. Barnabei, P. Pupillo,
F. Roversi Monaco. Published by the University of Bologna and its
Academy of Sciences, Bologna, 1998, Galvani celebrations.
\item{15.} M.A. Bouchiat and C.~Bouchiat, {\it J.~Phys.} (Paris) {\bf 35} (1974) 899, {\it ibid. \bf 36}(1975) 493;\br
C.S.~Wood, {\it et al, Science \bf 257} (1997) 1759; \br S.C.~Bennet and C.E.~Wieman, {\it Measurement
of the $6S \ra 7S$ transition polarizability in atomic cesium and an improved test of the standard model},
arXiv:hep-ex/9903022.
\item{16.}M.~ Jamin, {\it Theoretical status of $\epsilon'/\epsilon$}, hep-ph/9911390.
\item{17.} A.~Zichichi, {]it Antimatter}, in Proveedings of the
8th International Wordkshop on ``Neutrino Telescopes", 23-26
February 1999, Venice, Italy.
\item{18.}N. Arkani-Hamed {\it et al}, {\it A Small Cosmological Constant from a Large Extra Dimension},
 hep-th/0001197.
\item{19.} Y.~Suzuki and Y.~Totsuka, {\it Nucl.~Phys. Proc.~Suppl \bf 77} (1999) 546 P;
L.~Maiani, {\it Cold Dark Matter and Massive Neutrinos, in the
Universe and in the Lab}. Talk presented at the Symposium ``Future
of Universe, Future of Earth, Future of Civilization", Budapest -
Debrecen, Hungary, 2-6 July 1999 (CERN preprint).
\item{20.}C.W.~Misner, K.S.~Thorne and J.A.~Wheeler, {\it Gravitation},  Freeman,  San
Francisco, 1973;\br S. Chandrasekhar, {\it The Mathematical
Theory  of Black Holes}, Clarendon Press, Oxford University Press;
\br K.S.~Thorne, {\it Black Holes: the Membrane Paradigm}, Yale
Univ. Press, New Haven, 1986.
 \item{21.} R.C.~Tolman, {\it Relativity, Thermodynamics and Cosmology} (Clarendon, Oxford, 1934), pp. 242-243;
R.~Oppenheimer and G.~Volkoff, {\it Phys.~Rev. \bf 55} (1939) 374.
\item{22.} S.W. Hawking, {\it Commun. Math. Phys.} {\bf 43} (1975) 199.
\item{23.} G. 't Hooft, {\it Nucl. Phys.} {\bf B256} (1985) 727.
\item{24.} S.W. Hawking, {\it Phys.~Rev.} {\bf  D14} (1976) 2460.
\item{25.}G. 't Hooft, {\it Phys.~Scripta  \bf T15} (1987) 143; {\it id.},
  {\it J. Mod. Phys.} {\bf A11} (1996)  4623 (gr-qc/9607022).
\item{26.} G. 't Hooft, ``Dimensional Reduction in Quantum Gravity", Essay dedicated to
Abdus Salam, Utrecht preprint THU-93/26 (gr-qc/9310026); {\it
id.}, ``Black holes and the dimensionality of space-time", in {\it
Proceedings of the Symposium ``The Oskar Klein Centenary''}, 19-21
Sept. 1994, Stockholm, Sweden. Ed. U. Lindstr\"om, World
Scientific 1995, p. 122; \br L. Susskind, L. Thorlacius and J.
Uglum, {\it Phys. Rev.} {\bf D48} (1993) 3743 (hep-th 9306069).

\bye